# Nanoconstriction spin-Hall oscillator with perpendicular magnetic anisotropy

B. Divinskiy[1*], V. E. Demidov[1], A. Kozhanov[2,3], A.B. Rinkevich[4], S. O. Demokritov[1,4], and S. Urazhdin[5]

[1]*Institute for Applied Physics and Center for Nanotechnology, University of Muenster, Corrensstrasse 2-4, 48149 Muenster, Germany*

[2]*Department of Physics and Astronomy, Georgia State University, Atlanta, GA 30303, USA*

[3]*National Research Nuclear University MEPhI, 115409 Moscow, Russia*

[4]*Institute of Metal Physics, Ural Division of RAS, Yekaterinburg 620041, Russia*

[5]*Department of Physics, Emory University, Atlanta, GA 30322, USA*

We experimentally study spin-Hall nano-oscillators based on [Co/Ni] multilayers with perpendicular magnetic anisotropy. We show that these devices are capable of single-frequency auto-oscillations at current densities comparable to those in the in-plane magnetized oscillators. The demonstrated oscillators exhibit large magnetization precession amplitudes, and their oscillation frequency is highly tunable by the electric current. These features make them promising for applications in high-speed integrated microwave circuits.

After the recent demonstration of magnetic nano-oscillators driven by pure spin currents,[1-2] these devices attracted significant attention,[3-13] since they can enable the implementation of efficient tunable nano-scale sources of microwave signals[3-5,11-13] and propagating spin waves,[8,10] characterized by moderate heat generation, high oscillation coherence, and flexible layout. Additionally, by utilizing pure spin currents not accompanied by charge currents, it became possible to achieve spin torque-driven excitations in magnetic insulators.[14-17]

The unique flexibility of spin current-driven oscillators has facilitated the development of a variety of device layouts including nano-gap,[1] nanoconstriction,[4,11] nano-wire,[5] and nano-contact[7] geometries. Among these geometries, the nanoconstriction oscillators driven by the spin-Hall effect (SHE)[18,19] are particularly attractive, thanks to the simplicity of fabrication, the ease of electronic and optical characterization, and good oscillation characteristics at room temperature. It was recently shown that the spectral coherence and the output power of microwave generation by nanoconstriction oscillators can be significantly improved by magnetizing them out of the plane of the active magnetic layer.[11] In contrast to the in-plane magnetized configuration, where the auto-oscillations evolve from the modes strongly localized at the edges of the nanoconstriction region, in the out-of-plane geometry, the auto-oscillations occupy a significantly larger magnetic volume,[20] resulting in a higher stability of the oscillation, and providing the possibility for mutual synchronization of neighboring devices.[13] In addition, static magnetic field applied out of the plane of the magnetic layer can lift the degeneracy of the spin wave spectrum at the oscillation frequency. Such degeneracy is known to facilitate the onset of nonlinear magnon scattering, which limits the oscillation amplitude in the in-plane magnetized devices.[5,12]

A clear challenge for the out-of-plane magnetized geometry is associated with the need for large magnetic fields to overcome the demagnetizing field of the magnetic film. This challenge can be addressed by utilizing magnetic materials with perpendicular magnetic anisotropy (PMA). PMA materials have been intensively explored for applications in traditional spin-torque nano-oscillators operating with spin-polarized electric currents.[9] However, their suitability for the implementation of devices driven by pure spin currents, and capable of stable coherent oscillations at room temperature, remains largely untested.[21]

In this Letter, we demonstrate a nanoconstriction spin-Hall oscillator based on the Pt/[Co/Ni] multilayer with PMA. We show that this system exhibits single-frequency spin current-driven auto-oscillations at moderate electrical current densities in the Pt layer, comparable to those typical for the spin-Hall oscillators based on the in-plane magnetized Permalloy films. An important benefit of the oscillators based on the PMA materials is that they do not support spin waves spectrally degenerate with the oscillation, enabling large magnetization precession amplitudes not limited by the nonlinear magnon scattering processes. In the absence of these limiting phenomena, a very high tunability of the auto-oscillation frequency by electric current becomes possible, reaching values over 1 GHz/mA in our measurements.

Figure 1(a) shows the layout of the tested devices. They are based on a Pt(5)/[Co(0.2)/Ni(0.8)]$_4$/Ta(3) magnetic multilayer with PMA. Here, thicknesses are in nanometers. The multilayer is patterned into the shape of a bow-tie nanoconstriction with the width of 100 nm, the opening angle of 22°, and the radius of curvature of about 50 nm. The dc electric current $I$ flowing in the plane of the multilayer is converted by the SHE in Pt into an out-of-plane spin current $I_S$. The spin current injected into the [Co/Ni] multilayer exerts spin-transfer torque (STT)[22,23] on its magnetization $M$. Depending on the relative orientation of the polarization of the spin current and the magnetization $M$, the STT either increases or reduces the effective magnetic damping in the [Co/Ni] multilayer.[24] When the damping is completely compensated, the magnetization exhibits steady-state auto-oscillations.[12] The abrupt narrowing of the Pt layer in the nanoconstriction region results in a strong local increase of the density of the electric current, and consequently of the spin current, allowing one to achieve auto-oscillations in the active device area at moderate driving currents. Based on the symmetry of the SHE, the polarization of the spin current is parallel to the *x*-direction as defined in Fig. 1(a). Under these conditions, the effects of STT vanish if *M* is exactly perpendicular to the plane of the magnetic multilayer. To achieve spin current-induced oscillation, we apply an in-plane static magnetic field $H_\parallel$ < 2000 Oe (see Fig. 1(a)), which tilts the magnetization towards the *x*-axis. Additionally, to prevent the switching of the magnetization by SHE, we apply a small out-of-plane magnetic field $H_\perp$ = 200 Oe.

To characterize the static and the dynamic magnetic properties of the [Co/Ni] multilayer under the conditions described above, we performed micromagnetic simulations using the MUMAX3 software.[25] The parameters of the multilayer used in the calculations – the saturation magnetization $M_s$ = 490 kA/m and the out-of-plane magnetic anisotropy $K_u$ = 0.23 MJ/m$^3$ – were determined from independent measurements utilizing vibrating-sample and magneto-optical Kerr-effect magnetometries. As shown in Fig. 1(b), increasing the in-plane field $H_\parallel$ from 0 to 2000 Oe results in the reduction of the magnetization angle $\theta$ measured relative to the *x*-axis from 90º to 57º, while the frequency of the uniform ferromagnetic resonance (FMR) in the [Co/Ni] multilayer decreases from 9.5 GHz to about 8 GHz. We note that, despite significant tilting of the magnetization, the form of the dispersion relation of spin waves remains similar to that for the out-of-plane magnetized films[26] even at the largest applied in-plane field $H_\parallel$=2000 Oe, Fig. 1(c). The dispersion is nearly isotropic in the film plane. In contrast to films with in-plane magnetization, there are no finite-*k* spin wave modes degenerate with the uniform *k*=0 mode. This is particularly important for the spin current-induced auto-oscillation, since spin-wave degeneracy is known to cause nonlinear magnon scattering, limiting the achievable oscillation amplitude.[5,12]

We detect the spin current-induced magnetization dynamics by using micro-focus Brillouin light scattering (BLS)[27] spectroscopy. We focus the probing laser light with the wavelength of 532 nm into a diffraction-limited spot on the surface of the [Co/Ni] multilayer (Fig. 1(a)) and analyze the spectrum of light inelastically scattered from magnetic excitations. The detected BLS signal is proportional to the intensity of the magnetic oscillations at the location of the probing light spot.

Figure 2(a) shows the BLS spectra obtained at $H_\parallel$ = 2000 Oe, with the probing spot positioned at the center of the nanoconstriction. The onset of auto-oscillations is signified by the emergence of an intense spectral peak at $I$ = 2.3 mA. The amplitude of the auto-oscillation peak rapidly increases with increasing current, reaches a maximum at $I$ = 3.1 mA, and then the peak starts to broaden while its amplitude decreases. The evolution of the auto-oscillation characteristics with current is analyzed in Fig. 2(b). As seen from these data, the peak intensity (point-down triangles) increases at currents up to $I$=3.1 mA, but starts to decease at larger currents. Meanwhile, the intensity integrated over the BLS spectrum, which reflects the energy stored in the auto-oscillation mode, grows linearly at the same rate at currents both below and above 3.1 mA (point-up triangles in Fig. 2(b)).

We emphasize that these behaviors are qualitatively different from those observed in the in-plane magnetized spin-current oscillators,[12] where both the peak amplitude and the integral intensity decrease at large currents. The decrease of integral intensity is usually associated with the onset of nonlinear magnon scattering, resulting in energy transfer from the auto-oscillation mode into incoherent short-wavelength spin waves whose frequencies are degenerate with the auto-oscillation. In devices with PMA of active layer, the spin wave dispersion relation exhibits a minimum at $k$=0 (see Fig. 1(c)). Consequently, there are no short-wavelength spin wave modes that are degenerate with the quasi-uniform auto-oscillation mode. Thus, the nonlinear magnon scattering mechanism, which reduces the energy of the auto-oscillation mode driven by the spin current, is absent, and the integral intensity of auto-oscillations continues to grow even at large currents. The broadening of the auto-oscillation peak at larger currents, and the associated reduction of the peak amplitude, can be attributed to the onset of strongly anharmonic large-amplitude magnetization precession. We emphasize that this regime cannot be achieved in the in-plane magnetized devices, because the nonlinear magnon scattering mechanism described above limits the precession amplitude.

Our conclusion that the auto-oscillation in the studied devices reaches very large amplitudes is supported by the analysis of the dependence of the spectral peak frequency on the driving current, Fig. 2(c). Thanks to the high sensitivity of BLS, the spectral peak can be detected even at currents below the onset of auto-oscillation at $I_C$=2.3 mA, although its amplitude remains small (see Fig. 2(b)). In this regime, spin current enhances incoherent magnetization fluctuations in all the dynamic magnetic modes,[28] which leads to the gradual reduction of the effective magnetization, and results in the decrease of the spectral peak frequency. Continuous evolution of the peak frequency through the onset of auto-oscillation shows that the auto-oscillation mode evolves from the FMR mode of the [Co/Ni] multilayer.

At the onset of the auto-oscillation, the energy supplied by the spin current becomes channeled predominantly into the auto-oscillation mode,[12] resulting in the fast growth of its intensity, and an accompanying nonlinear frequency shift of auto-oscillations. While both the reduction of the effective magnetization and the nonlinear frequency shift lead to the decrease of the spectral peak frequency, the data in Fig. 2(c) exhibit a clear kink at $I_C$=2.3, signifying a transition between the two mechanisms.

As seen from Fig. 2(c), the auto-oscillation frequency decreases by about 1 GHz when the driving current is increased from $I_C$=2.3 mA to 3.5 mA. This decrease is about an order of magnitude larger than the current-dependent frequency variation observed in SHE devices with in-plane magnetization.[12] Since this frequency variation in the PMA device originates from the increase of the magnetization precession amplitude, one can estimate the corresponding precession angle. Our micromagnetic simulations show that to achieve nonlinear frequency shift of 1 GHz, the precession angle must increase by about 20º. We point out that the observed strong variation of the auto-oscillation frequency with current is advantageous for applications requiring a large tunability of the frequency of the generated microwave signals.

Figure 3(a) shows the dependence of the auto-oscillation frequency $f_A$ on the static field measured at $I=I_C$, as well as $df_A/dI$ reflecting the frequency tunability by the current. The dependence of $f_A$ on field is moderate, spanning a range of about 2 GHz when $H_\parallel$ is varied between 1000 and 2000 Oe. In contrast, the value of $df_A/dI$ remains very large over the entire studied range of fields, exceeding 1 GHz/mA at $H_\parallel$ = 1000 Oe. We note that from the practical standpoint, frequency tuning by current is simpler to implement and much faster than tuning by the static magnetic field. Thus, SHE oscillators with PMA can be particularly promising for the implementation of high-speed integrated microwave circuits.

The possibility of frequency tuning by varying the field is somewhat compromised by the relatively strong dependence of the density of the critical current $J_C$ on the static field (Fig. 3(b)). It originates from the strong field dependence of the angle $\theta$ between the equilibrium magnetization direction and the direction of spin current polarization (Fig. 1(b)). Indeed, the theory of STT predicts that the efficiency of spin current-induced excitation is proportional to $\cos(\theta)$.[29] Therefore, increase of $\theta$ with decreasing $H_\parallel$ leads to a larger critical current density. Nevertheless, despite the relatively large magnetic damping typical for [Co/Ni] multilayers,[30] $J_C$ remains comparable to that for the in-plane magnetized Permalloy-based SHE nano-oscillators[12] over the entire studied field range $H_\parallel$ = 1000 - 2000 Oe.

Finally, we analyze the nature of the auto-oscillation mode. By scanning the probing laser spot over a 0.7 μm by 0.7 μm area centered at the nanoconstriction, we obtain a two-dimensional map of the intensity of the dynamic magnetization shown in Fig. 4(a). The map shows that the large-intensity oscillation region is approximately circular, and has a full width at half maximum of about 350 nm. Considering that the measured map represents a convolution of the actual distribution of the dynamic magnetization with the probing laser spot with the diameter of about 250 nm, we conclude that the auto-oscillations are localized in a relatively large area with characteristic dimensions of about 250 nm. To check whether this localization is associated with the existence of the edge modes,[20] we analyze the calculated spatial distribution of the static internal field $H_{int}$ in the structured [Co/Ni] multilayer (Fig. 4(b)). Similarly to the in-plane magnetized nanoconstrictions, the internal field is noticeably reduced in the vicinity of the edges. However, since the FMR frequency in the multilayer with PMA increases with the decrease of the in-plane static magnetic field (Fig. 1(b)), the reduction of the internal field at the edge cannot lead to the formation of localized edge modes. For the same reason, the localization cannot be associated with the influence of the Oersted field of the driving current. Therefore, localization of auto-oscillations in the studied system is likely caused by the reduction of the effective magnetization due to the injection of spin current. Indeed, this reduction leads to a significant decrease of the FMR frequency in the nanoconstriction region even at $I < I_C$ (Fig. 2(c)). This effect can be also described as the formation of an effective local potential well for spin waves.[7] We note that this localization mechanism results in the large spatial dimensions of the auto-oscillation mode determined by the geometry of spin injection region, consistent with our experimental observations. It is favorable for applications, since the large volume of auto-oscillation region reduces the sensitivity to thermal fluctuations and facilitates the implementation of systems comprised of interacting nano-oscillators.[12,13]

In conclusion, we have demonstrated spin-Hall nano-oscillators based on magnetic multilayers with PMA. Our results show that PMA materials can provide a significant improvement of the characteristics of spin-current oscillators. They can eliminate the nonlinear mechanisms limiting the oscillation amplitude, and enable a large tunability of the auto-oscillation frequency by the dc current. Our observations should stimulate further developments in the field of nano-oscillators driven by pure spin currents and their applications in integrated microwave circuits.

This work was supported in part by the Deutsche Forschungsgemeinschaft, the NSF Grant Nos. ECCS-1509794 and DMR-1504449, and the program Megagrant № 14.Z50.31.0025 of the Russian Ministry of Education and Science.

**References**

1. V. E. Demidov, S. Urazhdin, H. Ulrichs, V. Tiberkevich, A. Slavin, D. Baither, G. Schmitz, S. O. Demokritov, Nature Mater. **11**, 1028–1031 (2012).

2. L. Liu, C.-F. Pai, D. C. Ralph, and R. A. Buhrman, Phys.Rev. Lett. **109**, 186602 (2012).

3. R. H. Liu, W. L. Lim, and S. Urazhdin, Phys. Rev. Lett. **110**, 147601 (2013).

4. V.E. Demidov, S. Urazhdin, A. Zholud, A.V. Sadovnikov, S.O. Demokritov, Appl. Phys. Lett. **105**, 172410 (2014).

5. Z. Duan, A. Smith, L. Yang, B. Youngblood, J. Lindner, V. E. Demidov, S. O. Demokritov, I. N. Krivorotov, Nature Commun. **5**, 5616 (2014).

6. M. Ranjbar, P. Durrenfeld, M. Haidar, E. Iacocca, M. Balinskiy, T. Q. Le, M. Fazlali, A. Houshang, A. A. Awad, R. K. Dumas, and J. Akerman, IEEE Magn. Lett. **5**, 3000504 (2014).

7. V. E. Demidov, S. Urazhdin, A. Zholud, A. V. Sadovnikov, A. N. Slavin, S. O. Demokritov, Sci. Rep. **5**, 8578 (2015).

8. V. E. Demidov, S. Urazhdin, R. Liu, B. Divinskiy, A. Telegin, S. O. Demokritov, Nature Commun., **7**, 10446 (2016).

9. T. Chen, R. K. Dumas, A. Eklund, P. K. Muduli, A. Houshang, A. A. Awad, P. Dürrenfeld, B. G. Malm, A. Rusu, J. Åkerman, Proc. IEEE **104**, 1919–1945 (2016).

10. B. Divinskiy, V. E. Demidov, S. O. Demokritov, A. B. Rinkevich, S. Urazhdin, Appl. Phys. Lett., **10**9, 252401 (2016).

11. P. Dürrenfeld, A. A. Awad, A. Houshang, R. K. Dumas, and J. Åkerman, Nanoscale **9**, 1285 (2017).

12. V. E. Demidov, S. Urazhdin, G. de Loubens, O. Klein, V. Cros, A. Anane and S. O. Demokritov, Phys. Rep. **673**, 1-31 (2017).

13. A. A. Awad, P. Dürrenfeld, A. Houshang, M. Dvornik, E. Iacocca, R. K. Dumas, and J. Åkerman, Nature Physics **13**, 292–299 (2017).

14. M. Collet, X. de Milly, O. d'Allivy Kelly, V. V. Naletov, R. Bernard, P. Bortolotti, J. Ben Youssef, V.E. Demidov, S. O. Demokritov, J. L. Prieto, M. Munoz, V. Cros, A. Anane, G. de Loubens, O. Klein, Nature Commun. **7**, 10377 (2016).

15. V. E. Demidov, M. Evelt, V. Bessonov, S. O. Demokritov, J. L. Prieto, M. Muñoz, J. Ben Youssef, V.V. Naletov, G. de Loubens, O. Klein, M. Collet, P. Bortolotti, V. Cros, A. Anane, Sci. Rep. **6**, 32781 (2016).

16. C. Safranski, I. Barsukov, H. K. Lee, T. Schneider, A. Jara, A. Smith, H. Chang, K. Lenz, J. Lindner, Y. Tserkovnyak, M. Wu, I. Krivorotov, Spin caloritronic nano-oscillator, arXiv:1611.00887 (2016).

17. V. Lauer, M. Schneider, T. Meyer, C. Dubs, P. Pirro, T. Brächer, F. Heussner, B. Lägel, V. I. Vasyuchka, A. A. Serga, B. Hillebrands, A. V. Chumak, Auto-oscillations in YIG/Pt microstructures driven by the spin Seebeck effect, arXiv:1612.07305 (2016).

18. M. I. Dyakonov, V. I. Perel, Sov. Phys. JETP Lett. **13**, 467–469 (1971).

19. J. E. Hirsch, Phys. Rev. Lett. **83**, 1834–1837 (1999).

20. M. Dvornik, A. A. Awad, J. Åkerman, On the origin of magnetization auto-oscillations in constriction-based spin Hall nano-oscillators, arXiv:1702.04155 (2017).

21. R. H. Liu, W. L. Lim, and S. Urazhdin, Phys. Rev. Lett. **114**, 137201 (2015).

22. J.C. Slonczewski, J. Magn. Magn. Mater. **159**, L1–L7 (1996).

23. L. Berger, Phys. Rev. B **54**, 9353–9358 (1996).

24. K. Ando, S. Takahashi, K. Harii, K. Sasage, J. Ieda, S. Maekawa, and E. Saitoh, Phys. Rev. Lett. **101**, 036601 (2008).

25. A. Vansteenkiste, J. Leliaert, M. Dvornik, M. Helsen, F. Garcia-Sanchez, and B. Van Waeyenberge, AIP Advances **4**, 107133 (2014).

26. A. G. Gurevich and G. A. Melkov, Magnetization Oscillation and Waves, CRC Press, Boca Raton (1996).

27. V. E. Demidov and S. O. Demokritov, IEEE Trans. Mag. **51**, 0800215 (2015).

28. V. E. Demidov, S. Urazhdin, E. R. J. Edwards, M. D. Stiles, R. D. McMichael, and S. O. Demokritov, Phys. Rev. Lett. **107**, 107204 (2011).

29. A. N. Slavin and P. Kabos, IEEE Trans. Magn. **41**, 1264 (2005).

30. J.-M. L. Beaujour, W. Chen, K. Krycka, C.-C. Kao, J. Z. Sun, and A. D. Kent, Eur. Phys. J. B **59**, 475–483 (2007).

**Figure captions**

Fig. 1 (color online) (a) Schematic of the experiment. (b) Calculated static-field dependence of the out-of-plane angle of the equilibrium magnetization (circles) and of the uniform FMR frequency (squares). Curves are guides for the eye. (c) Calculated spectrum of spin waves in the [Co/Ni] multilayer magnetized by the in-plane field of 2000 Oe.

Fig. 2 (color online) (a) Representative BLS spectra of magnetization auto-oscillations recorded at different driving currents, as labeled. Symbols – experimental data, lines – fit by the Lorentzian function. (b) Peak intensity (point-down triangles) and intensity integrated over the BLS spectrum (point-up triangles) vs dc current. (c) Current dependence of the central frequency of the detected spectral peak. Symbols – experimental data, curve – a guide for the eye. Arrow marks the critical current $I_C$ corresponding to the onset of auto-oscillations. The data were obtained at $H_\parallel$ = 2000 Oe.

Fig. 3 (color online) (a) Magnetic-field dependence of the auto-oscillation frequency $f_A$ measured at $I=I_C$ (squares) and of the tunability of the auto-oscillation frequency by the current (circles). (b) Magnetic-field dependence of the density of the critical current. Symbols – experimental data, curves – guides for the eye.

Fig. 4 (color online) (a) Normalized color-coded spatial map of the dynamic magnetization in the auto-oscillation mode recorded at $I$ = 3 mA and $H_\parallel$ = 2000 Oe. (b) Spatial distribution of the static internal in-plane magnetic field $H_{int}$ in the [Co/Ni] multilayer calculated at $H_\parallel$ = 2000 Oe.

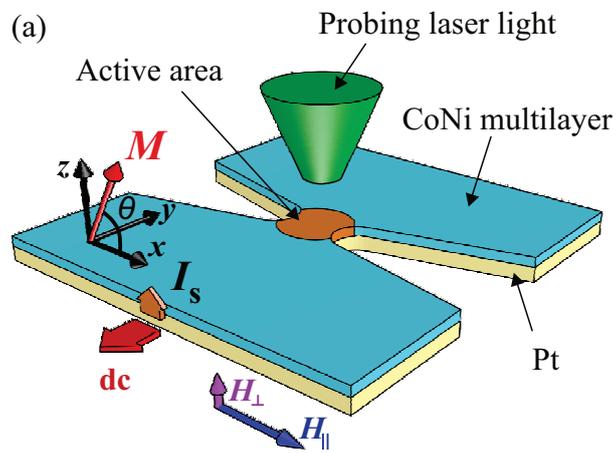

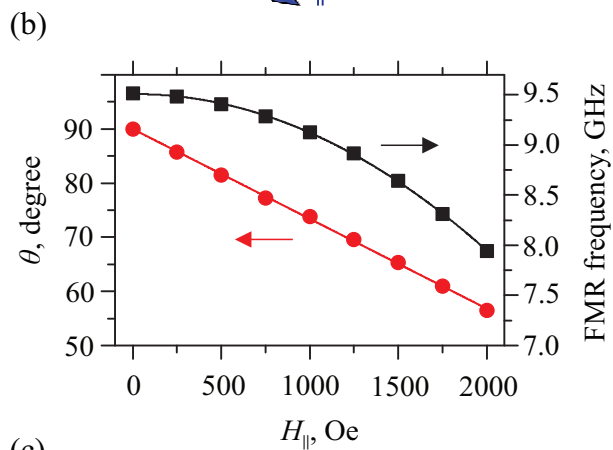

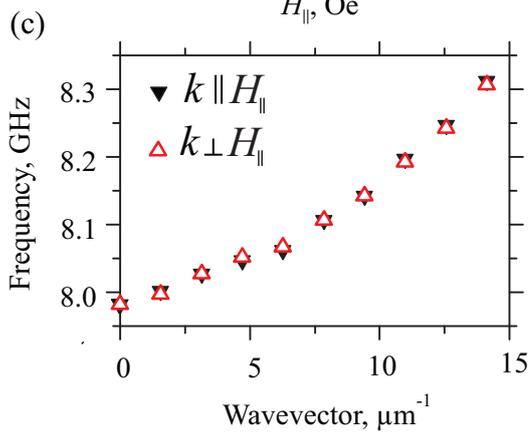

Fig. 1

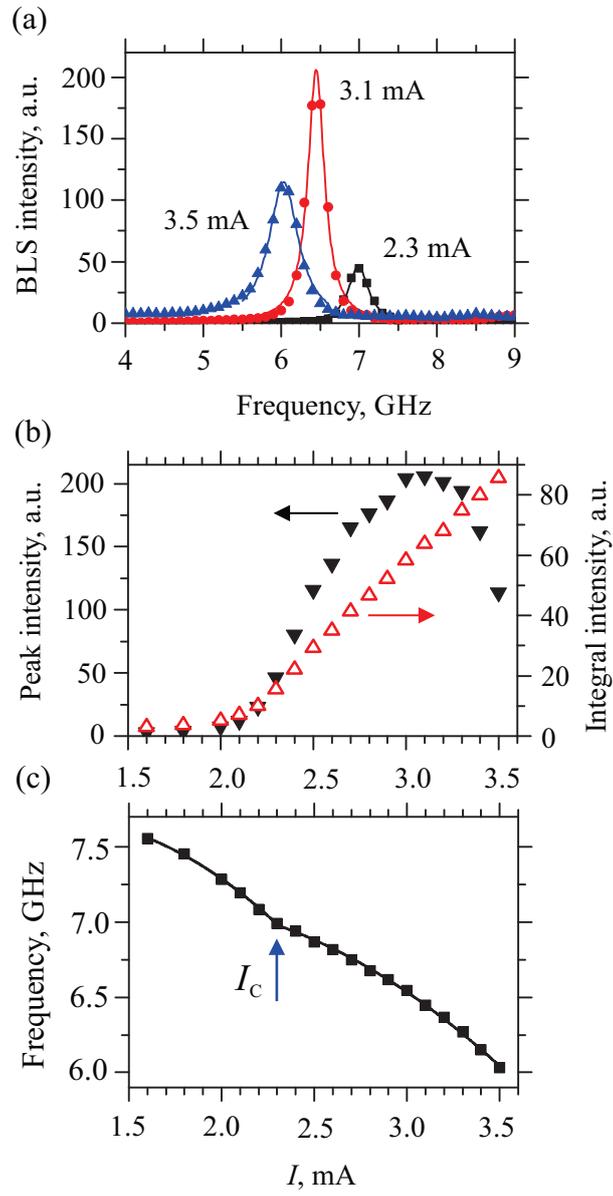

Fig. 2

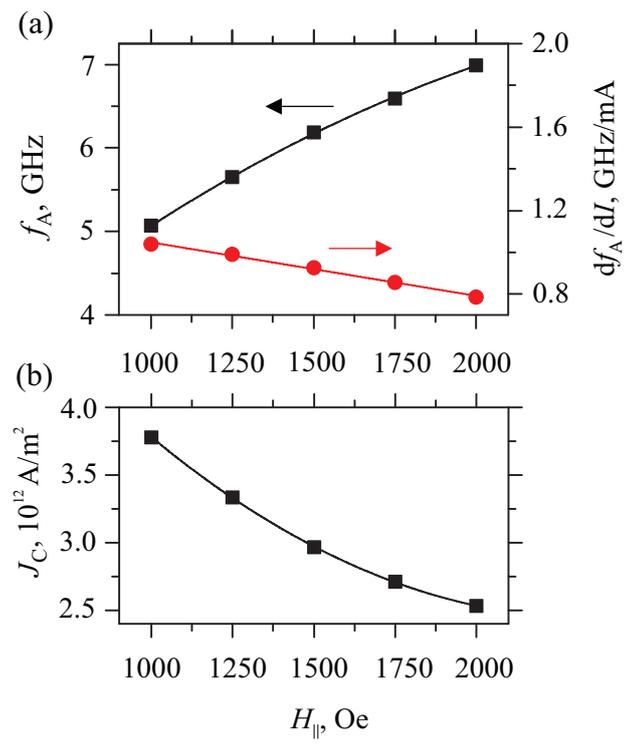

Fig. 3

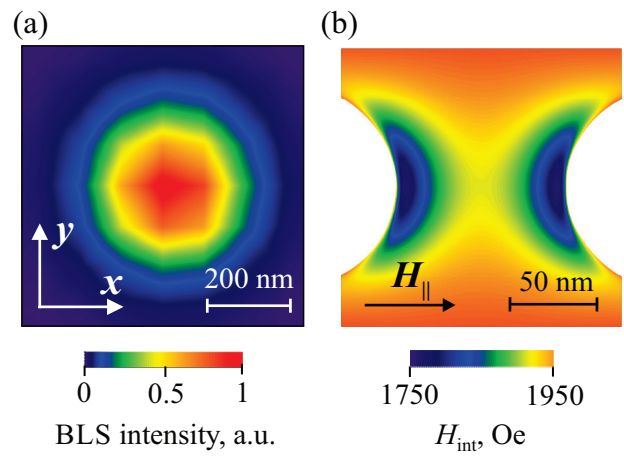

Fig. 4